



\def\rarr{\rightarrow}

\def\bar#1{\overline{#1}}
\def\bold#1{\setbox0=\hbox{$#1$}%
     \kern-.025em\copy0\kern-\wd0
     \kern.05em\copy0\kern-\wd0
     \kern-.025em\raise.0433em\box0 }

\def\dslash{\not{\hbox{\kern-2pt $\partial$}}}
\def\Dslash{\not{\hbox{\kern-4pt $D$}}}
\def\Qslash{\not{\hbox{\kern-4pt $Q$}}}
\def\pslash{\not{\hbox{\kern-2.3pt $p$}}}
\def\kslash{\not{\hbox{\kern-2.3pt $k$}}}
\def\qslash{\not{\hbox{\kern-2.3pt $q$}}}
\def\pairof#1{#1^+ #1^-}

\def\ee{\pairof{e}}
\def\CM{{\rm CM}}
\def\ECM{E_\CM}
\def\BR{{\rm BR}}

\def\L{{\cal L}}
\def\H{{\cal H}}
\def\M{{\cal M}}

\def\sstw{\sin^2\theta_w}
\def\cstw{\cos^2\theta_w}

\def\msb{{\bar{\ssstyle M \kern -1pt S}}}

\def\Re{{\rm Re}}
\def\Im{{\rm Im}}
\def\refmark#1{~[#1]}
\def\etal{{\it et.~al.}}

\def\roots{$\sqrt{s}$}

\def\lunit{cm$^{-2}$sec$^{-1}$}


%
\pubnum{6582}
\pubtype{(T/E)}
\date{August, 1994}
\titlepage
\title{Complementarity of $\ee$ and $pp$ Colliders \break
          for the Exploration of Electroweak Symmetry Breaking}
\author{Michael E. Peskin\doeack}
\SLAC
\vfil
\centerline{Invited lecture presented at the 22nd INS Symposium}
\centerline{Physics with High Energy Colliders}
\centerline{Tokyo University, March 8--10, 1994}
\abstract{I review the physics capabilities of the machines proposed
     for the next generation of high-energy experimentation: in hadron
          physics, the LHC, and in electron physics, a $500$--$1500$
         GeV
       $\ee$ linear
        collider.  Using for illustration two specific models of
         electroweak symmetry breaking, I show how the $pp$ and
            $\ee$ techniques are expected to complement one another
           in the exploration of the next scale of physics.}

\endpage
\sequentialequations
\singlespace
\chapter{Introduction}

When we look forward to the future of elementary particle
 physics, we anticipate
discoveries of new particles and phenomena at increasingly higher
energies.  To plan for this future, we need to design and
construct accelerators which access these new energy regimes.
It is well appreciated, though, that this planning raises many
difficulties.
New facilities for high-energy physics
are extremely complex, requiring a decade or more for their
planning and construction.  They are also extremely expensive, so
much so that the high energy accelerators of the future will require
funding through international collaboration.  Thus, we must
evolve a persuasive plan that is scientifically grounded and can win
wide acceptance.

 Our situation would be made easier if there were a single ideal
machine design which would allow us to answer our most pressing
questions.  Over the history of our field, there have been many
claims that a particular machine configuration would provide the
essential clues that we are seeking.  In the United States, both the
SLAC and Fermilab accelerators were proposed by men with intense,
and completely divergent,
personal visions of the correct next step in high-energy physics.
More recently, we have seen the SSC put forward in the United States
 as the crucial
accelerator for the future, only to leave a vacuum in our
national planning when the SSC project was cancelled last year.
Still,
historically, however powerful the claims and even the achievements
of a particular technique have been, our understanding of physics
has grown through
 the synthesis of experimental information from many
sources.  Thus, as we evolve our global plan, this should include
      different types of experimental facilities which complement
one another.

\REF\LHCD{LHC Study Group,
\sl Design Study of the Large Hadron Collider.
                   \rm (CERN, 1991).}
\REF\NLCdes{G. Loew, in \sl Proceedings of the ECFA Workshop on
          $\ee$ Linear Colliders (LC92), \rm R. Settles, ed.
                 (MPI Munich, 1993).}
\REF\TechNLC{B. Wiik, in \sl  Physics and Experiments with Linear
         $\ee$ Colliders, \rm vol. I, F. A. Harris, S. L. Olsen, S.
          Pakvasa, and X. Tata, eds.  (World Scientific, Singapore,
                  1993).}

  The most technically promising means to achieve the next step
in center of mass energy are through proton-proton and electron-positron
colliders.  On the proton-proton side, the next step is almost
assured through the fact that CERN has made the Large Hadron Collider
(LHC) its highest priority project\refmark\LHCD.
   This project should give us access
to parton-parton collisions at multi-TeV energies.
 On the electron-positron side, the
situation is more uncertain.  The technical feasibility of the
next collider is now clear, with prototypes of all of the major
components
now completed or under construction at laboratories
around the world.    There are now several
competing designs for machines that would begin operation at
400-500 GeV and would be expandable in energy up to about
 1.5 TeV\refmark{\NLCdes,\TechNLC}.
 However, it is
much less clear, at least to the broader world community, what role
the next $\ee$ collider would play in relation to the high-energy
experiments being done on the proton-proton side.

  It is this question that I would like to address:
How will experiments at the next $\ee$ collider complement experiments
to be done at the next $pp$ collider, and what $\ee$
center-of-mass energy is
needed to achieve the best match between these facilities?

  In discussions among physicists, and even in the literature, one often
sees facile and oversimplified answers to these questions.  It is argued,
for example, that
 $pp$ and $\ee$ facilities are complementary only
when they have comparable parton center-of-mass energy, or that proton
machines are best for `discovery' while $\ee$ facilities are best for
`precision studies'.
The history of the discovery of ingredients of the
standard model  of course gives a more complicated picture.  For example,
the gluon was discovered at an $\ee$ collider, while quantitative
information on its interactions was derived from $p\bar p$ experiments.
Thus, we must check our preconceptions against
                  detailed analysis.  A purely historical approach,
however, may not extrapolate simply to the
problems of future experimentation.

In this lecture, I will take a different
approach which concentrates on a general issue of great importance
to the future colliders.  Elementary particle physicists have now
established the standard model gauge theory of strong, weak, and
electromagnetic interactions.  This theory has met stringent quantitative
tests, particularly in the sector of high-energy weak interactions.
But the
 theory has an obvious difficulty:  It requires that the
weak interaction gauge symmetry $SU(2)\times U(1)$ be spontaneously
broken, but it does not provide a physical mechanism for this
symmetry breaking.   One cannot approach most of the remaining
mysteries of particle physics---in particular, the questions of the
nature and mass spectrum of the quarks and leptons---without
understanding the solution to this problem.  Since the
mass scale of electroweak symmetry breaking is the $W$ mass scale,
 the solution to this problem should soon be experimentally
acccessible.  This problem was given as the main justification for
building the SSC, and it must figure strongly in the motivation for
any other future collider.  Indeed, because of the successes of the
standard electroweak model in its confrontation with experiment,
the impetus to solve this problem
is even greater now than it was ten years ago.

  Models of many different types have been proposed to solve the
problem of electroweak symmetry breaking. Every such model entails
new particles and forces.   At this moment, we have
very little information to discriminate these models.
But in the era of the next generation colliders, we should expect to
discover these new particles and begin the exploration of a new
sector of the fundamental interactions.

To analyze the requirements for this exploration,
 I will review the properties of two representative
models of electroweak symmetry breaking and
the experiments at future $pp$ and $\ee$ colliders that should give us
experimental insights into their structure.   Realistic
models of electroweak
symmetry breaking are complex   and multifaceted.  By analyzing the
variety of phenomena associated with a given model, we will gain an
appreciation of the richness of the phenomena which should be
 uncovered at the
next scale.  At the same time, we will obtain a concrete picture of the
complementary roles that  $pp$ and $\ee$ experiments would play in
the experimental elucidation of each model.  At the end of the lecture,
I will contrast the results of these explicit analyses with the
standard rules of thumb on the comparison of colliders.

\newpart{Two Models of Electroweak Symmetry Breaking}

  Since the main thrust of this lecture will involve the analysis of
theoretical models, I should begin by justifying my choice of models to
examine.   In the literature, one finds many models of electroweak
symmetry breaking with one or more Higgs doublet fields, more complex
particle multiplets, and possibly complicated strong-coupling
dynamics.  Which should we take as our examples?

\REF\GHP{L. Susskind, \sl Phys. Rev. \bf D20, \rm 2619 (1979).}
 The minimal model of electroweak symmetry breaking contains
a single Higgs field and gives rise to only one new particle, the
minimal Higgs boson.  Many authors have taken it as the basis for
detailed studies.  However, in my opinion, this model cannot be taken
seriously as a fundamental theory.  The model has well-known
pathologies:
These include the gauge hierarchy problem, the fact that the natural
value of the $W$ mass in this model is the grand unification or Planck
scale\refmark\GHP.
  In addition, in this model,
 all parameters of the quark and
lepton mass matrices are renormalizable coupling constants which must
be input to define the theory and which thus cannot be predicted.
These pathologies are an inevitable
 part of the minimal package; they
 characterize the fact that the
minimal Higgs model does not explain electroweak symmetry breaking
but, rather, is simply a parametrization of this phenomenon. In order to
explain electroweak symmetry breaking, we need to consider models which
contain richer dynamical possibilities.

\REF\Nilles{H. P. Nilles, \sl  Phys. Repts. \bf 110, \rm 1 (1984).}
\REF\HandK{H. E. Haber and G. L. Kane, \sl Phys. Repts. \bf 117, \rm 75
                 (1985).}
\REF\Lang{P. Langacker and M.-X. Luo, \sl Phys. Rev. \bf D44, \rm 817
                  (1991).}

   Models that present a mechanism for $SU(2)\times U(1)$ breaking
fall into one of two classes, depending on whether the Higgs boson is
taken to be elementary or composite.  If the model includes an
elementary Higgs field, it must contain some mechanism to cancel the
arbitrary additive mass renormalization of this scalar field.  The
only known mechanism to achieve this cancellation is supersymmetry.
The assumption of supersymmetry brings in a new sector of particles and
a complex array of new interactions.  However, it also brings some
important advantages:  Within supersymmetry, there is a natural
mechanism for $SU(2)\times U(1)$  breaking, since the Higgs field
which couples to the heavy top quark obtains a negative (mass)$^2$
renormalization.   The supersymmetric renormalization group equations
also naturally relate the values of the coupling constants obtained
at LEP to the predictions of a grand unified gauge theory.  These and
other features of supersymmetric models are reviewed in refs.
\Nilles--\Lang.

\REF\RMPTC{R. Kaul, \sl Rev. Mod. Phys. \bf 55, \rm 449 (1983).}
\REF\Appelq{K. Lane, in \sl Proceedings of the Theoretical Advanced
           Study Institute (TASI 93), \sl S. Raby, ed.
            (World Scientific, Singapore, 1994).}

  If the model does not include an elementary Higgs field, some new
strong interaction dyanmics must be provided to create the composite
state which acquires a vacuum expectation value.  The simplest way
to achieve this is by postulating a new strong interaction
gauge theory of fermions at a mass scale of about 1 TeV.  Then the
breaking of $SU(2)\times U(1)$ can procede by the same mechanism that
breaks chiral $SU(2)\times SU(2)$ in the familiar strong interactions.
In a theory constructed in this way, the new strong interaction is
called technicolor.  General aspects of technicolor models are
reviewed in refs. \RMPTC--\Appelq.

  In principle, there are many other ways in which new interactions at
the 1 TeV mass scale can induce the breakdown of $SU(2)\times U(1)$.
However, the two  examples of supersymmetry and technicolor
models have a particular advantage for the type of study that I will
describe here.  Since supersymmetry models involve only weak-coupling
dynamics, all relevant masses and cross sections can be computed from the
underlying parameters of the theory.  In technicolor models, one does
not have quite so much predictive power, but the properties of the
new strong interactions can be computed using phenomenological methods
borrowed from the study of the familiar strong interactions.  Thus,
for both types of model, there is a sizable literature on the
signatures of the new sector at future colliders.  We can make use of
this  literature
 to understand in detail the relation of $\ee$ and $pp$
experiments.

  I repeat that, in this lecture, I am not arguing that one of these
models must be correct.  It is quite likely that that solution of the
problem of electroweak symmetry breaking is more subtle, and that it
will require experimental elucidation.  What I am arguing is that we
should take known solutions to this problem seriously as
illustrative possibilities for
 the next scale in physics, and that we should pay attention to the
lessons they have to teach us.

  Both supersymmetry and
technicolor models are complex, and both provide a wide variety of
particles and phenomena that the new colliders should make visible.
It is not true in either model that
 a single discovery (for example, the sighting
of a Higgs boson) would clarify the physics.  Rather, this discovery
would be only the first step in a long and fascinating investigation.
In the models we have anticipated, we can work out in detail what tools
we will need for this investigation.   If Nature has chosen a model
that we have not
anticipated, we presumably will need even more experimental guidance.
 And even if these tools will be needed
only ten years from now, we must immediately set in motion
 the technical and political
processes that will make them available.

\newpart{`Discovery Reach'}

The simplest criterion for comparing the capabilities of $\ee$ and $pp$
machines is the parton-parton center of mass energy available to
produce new particles.  This criterion is often referred to as the
`energy' or `discovery reach' of a collider.
  I will argue later in this lecture that this
criterion is naive. Still, it is interesting to know, as a point of
reference, what this criterion predicts.

\REF\Snow{G. L. Kane and M. L. Perl, in
   \sl Elementary Particle Physics and Future Facilities \rm
           (Snowmass 1982),
             R. Donaldson, R. Gustafson, and F. Paige, eds.
                  (Fermilab, 1982).}
\REF\EHLQ{E. Eichten, I. Hinchliffe, K. Lane, and C. Quigg,
        \sl Rev. Mod. Phys. \bf 56, \rm 579 (1984).}
\REF\MyColl{M. E. Peskin, in \sl
  Physics in Collision 4, \rm A. Seiden, ed.  (\'Editions Fronti\`eres,
            Gif, 1984).}
\REF\Amaldi{U. Amaldi, in \sl Proceedings of the Workshop on Future
             Accelerators, \rm vol. I, J. H. Mulvey, ed. (CERN, 1987).}

  The usual way of making this comparison is to choose a sample list of
new physics processes, compare the energy needed to discover each
at a variety of colliders, and then form some sort of average.
Comparisons of this type can be found, for example, in refs.
 \Snow--\Amaldi.

\REF\MarkII{A. J. Weinstein, in \sl Proceedings of the 1989
              International Symposium on Lepton and Photon
            Interactions at High Energies, \rm M. Riordan, ed.
              (World Scientific, Singapore, 1990).}
 If the exotic particles under consideration have
electroweak quantum numbers, they can be pair-produced at $\ee$
colliders.  Typically, such particles can be produced for masses
almost up to \roots/2 for reasonable samples of integrated luminosity.
 As a striking example of this sensitivity,
one might consider the search limits reported by the Mark II experiment
at the SLC in ref. \MarkII, using a data sample of 500 $Z^0$ events.
In the remainder of this lecture, I will assume that future $\ee$
colliders will produce event samples of about 3000 events per year
per unit of R, comparable to the event samples of PEP and
PETRA.  This requires a luminosity increasing with energy according to
$$   \L \sim  10^{33} { \ECM^2\over (500 \ {\rm GeV})^2} \ \hbox{\lunit}
            \ .
  \eqn\Lguess$$
This  estimate \ is a factor 4--10 lower than
current design luminosities for the next generation linear collider.

  The discovery reach of a $pp$ collider is more difficult to
estimate. For any given new particle, one must find a signature which
can be observed in the $pp$ environment, define cuts which isolate
this signature from background, and then compute the number of $pp$
collisions required to produce a significant number of signal events
passing the cuts.  A comprehensive study of this kind was assembled
ten years ago
by Eichten, Hinchliffe, Lane, and Quigg in their review of
supercollider physics\refmark\EHLQ.  In the intervening time, many of the
analyses they presented have been made more sophisticated by inclusion of
the effects of hadronization and realistic  detectors, but the results of
their paper can still be used as a benchmark for broad comparisons.

\FIG\EHLQcompare{Comparison of the capabilities of $\ee$ and $pp$
colliders to   discover five representative signatures of new physics,
from ref.~\MyColl, assuming a $pp$ luminosity of $10^{33}$\lunit.
  The various signatures are:
a new $W$ boson, a heavy quark, a gluino, a heavy lepton, and a nonzero
compositeness scale $\Lambda$.  For $pp$ colliders, the discovery
reach was taken from ref. \EHLQ.  For $\ee$ colliders, the discovery
\roots\ values were taken to be
 $m_W$, $2m_Q$, $3m_{\widetilde g}$, $2m_L$,
$\Lambda/30$, respectively, for the five signatures.  The first of these
estimates
 assumes that there is also a new $Z$; the last reflects the
experience of PEP and PETRA.}
  In Fig. \EHLQcompare, I show a comparison of the
 discovery limits estimated in
ref. \EHLQ \ with those appropriate to $\ee$ colliders for five
particular new physics effects---a new $W$ boson, a heavy quark, a
gluino, a heavy lepton, and a nonzero scale of quark and lepton
compositeness\refmark{\MyColl}.
The comparable $\ee$ and $pp$
center of mass energies are given
for a  fixed $pp$ luminosity  of
$10^{33}$\lunit\ and an $\ee$ luminosity scaling according to \Lguess.
One can make similar figures for different assumptions about the
$pp$ luminosity.

   An average comparison is indicated in Fig. \EHLQcompare\ as the
dashed line.  This average corresponds to
$$  E_{\ee} \simeq 0.6 \bigl[ E_{pp} \bigr]^{1/2}  \bigl[\L_{pp}
             \bigr]^{1/6} \ ,
\eqn\scaling$$
where energies are in TeV and the proton-proton luminosity is in
units of $10^{33}$.  The dependence on luminosity realizes Kane's
rule of thumb\refmark\Snow:   a factor 2 in energy
is worth a factor 10 in luminosity.
  Putting in the parameters of the LHC (including
extrapolation    to $10^{34}$ luminosity), one finds an equivalent
$\ee$ center of mass energy of about 3 TeV.

   There are many reasons, however,
    why this comparison does not tell the full
story.  I have already noted that this sort of comparison concentrates
excessively on the first signal of new physics, and that it depends
on arbitrary assumptions about this signature.  But, more importantly,
this comparison ignores the fact that $\ee$ and $pp$ colliders
would typically observe different facets of a new  sector of
interactions.  In the models reviewed in the previous section, the
physics which leads to electroweak symmetry breaking is complex and
leads to new phenomena both in electroweak and in strong interaction
physics.  To understand the interrelation between the observables for
$\ee$ and $pp$ colliders, one must investigate the models in  more
detail.  In the next several sections, I will review the predictions of
supersymmetry and  technicolor models from this point of view.
We will see that these two complete disparate approaches to the problem
of electroweak symmetry breaking lead to similar conclusions about the
comparison of colliders, conclusions which are in both cases very
different from the simple scaling law of eq. \scaling.

\newpart{Supersymmetry: Higgs Sector}

  In this section and the next, I will compare the signatures of the
minimal supersymmetric standard model for $\ee$ and $pp$ colliders.
Supersymmetry is an example of a theory of electroweak symmetry breaking
in which all of the dynamics occurs as the result of weak-coupling
physics. For this reason, the consequences of the model are computable
in great detail, and a wide variety of signatures have been studied
quantitatively.

  If indeed Nature has chosen supersymmetry as the explanation for
electroweak symmetry breaking, the most important experimental issues
for the next generation of colliders will be the discovery of
the new particles present in this model, including the multiplet of
Higgs bosons, and the measurement of their couplings.  The most
important questions for $\ee$ and $pp$ colliders are summarized in
Table 1.  These questions divide into two parts.  First, supersymmetry
necessarily includes an extended Higgs boson sector which could, in
principle, be found in more general weakly-coupled models.  In this
section, I will discuss the study of this sector at future colliders.
Second, supersymmetry predicts a characteristic doubling of the
spectrum of elementary particles, with scalar partners of the quarks
and lepton and fermionic partners of the gauge bosons.  I will
discuss the study of these particles in the following section.

 In supersymmetric models, the same Higgs field cannot give mass to
both the $d$ and $u$ type quarks through Yukawa couplings invariant
under supersymmetry\refmark\Nilles.  Thus, supersymmetric models
require two Higgs doublet fields $\phi_1$, $\phi_2$. After the
components eaten by $W^\pm$ and $Z^0$ are removed, this sector
contains three neutral bosons  plus  the
charged pair $H^\pm$. The neutral bosons are derived from the
underlying fields by rotation through the mixing angles $\alpha$ and
$\beta$:
$$  \eqalign{
 CP\ {\rm even:} \qquad &  \pmatrix{\Re\, \phi_1 \cr \Re \, \phi_2\cr}
            \buildrel \alpha \over \longrightarrow
  \pmatrix{h^0 \cr H^0\cr}  \crr
 CP\ {\rm odd:} \qquad &  \pmatrix{\Im\, \phi_1 \cr \Im \,\phi_2\cr}
            \buildrel \beta \over \longrightarrow
  \pmatrix{\pi^0 \cr A^0\cr}\ .  \cr}
\eqn\Higgsbuild$$
The field $\pi^0$ is  eaten by the $Z^0$; the remaining three fields on
the right are associated with physical   particles.  The mixing angle
$\beta$ is related to the ratio of vacuum expectation values,
$$                \tan\beta = {\VEV{\phi_2}\over \VEV{\phi_1}} \ ;
\eqn\tanbetadef$$
this angle enters as a parameter throughout supersymmetry phenomenology.
In particular, when $\tan\beta$ is large, the $b$ quark
and $\tau$ lepton typically have large Yukawa couplings to the Higgs
sector to compensate the small vacuum expectation value which gives
these fermions mass.

 \REF\Haberrev{H. Haber, in \sl Perspectives on Higgs Physics, \rm
               G. L. Kane, ed.  (World Scientific, Singapore, 1992).}
\REF\Maianiandco{N. Cabbibo, L. Maiani, G. Parisi, and
R. Petronzio, \sl Nucl.
       Phys. \bf B158, \rm 295 (1979).}
\REF\Kaneandco{G. L. Kane, C. Kolda, and J. D. Wells,
       \sl Phys. Rev. Lett. \bf 70, \rm 2686 (1993).}
   In the most general Higgs theories, the mass spectrum of Higgs
bosons obeys only a few general constraints.  However,
the minimal supersymmetric
standard model provides some more specific
 relations among the masses of the Higgs
particles. Consider first the lightest Higgs boson $h^0$.
 At the tree level, the mass of this particle  obeys
$$  m(h^0) \leq m_Z\bigl| \cos 2\beta \bigr| \ .
\eqn\hzeromass$$
This bound is known to be  raised by large radiative corrections,
but still the upper limit is less than about 130 GeV\refmark\Haberrev.
In supersymmetric models with more complicated Higgs sectors, there is
no such simple formula.  Nevertheless,
the $h^0$ mass is still strongly restricted if the theory has a grand
unification: Under this assumption,
 the mass of the lightest
Higgs boson is controlled by a coupling constant which must be finite
at the grand unification scale and then is run by the renormalization
group to smaller values at lower mass scales\refmark\Maianiandco.
  One then finds an upper
limit of 200 GeV   on the mass of the lightest Higgs boson in the
broad class of grand unified models.
In nonminimal supersymmetric models, the lightest Higgs boson must lie
 below 150 GeV\refmark\Kaneandco.

    \REF\HHG{J. F. Gunion, H. E.
    Haber, G. Kane, and S. Dawson, \sl The Higgs Hunter's
               Guide. \rm (Addison Wesley, Redwood City, 1990).}
 For the heavier Higgs bosons, there are no similar upper bounds.
However, the masses of these bosons are linked together, in minimal
supersymmetry, by relations which read, at the tree level,
$$  \eqalign{
  m^2(H^+) = m^2 (A^0) + m_W^2 \cr
  m^2(H^0) = m^2 (A^0) + m_Z^2 - m^2(h^0)\ . \cr}
\eqn\masssqhh$$
It is convenient to consider the mass of the $A^0$ (henceforth, $m_A$)
as setting the scale of these heavier masses.  In models of
electroweak symmetry breaking through supersymmetry, $m_A$ is typically
in the range 200--500 GeV.  More details on the supersymmetric Higgs
boson spectrum can be found in refs. \HHG\ and \Haberrev.

\REF\KZ{Z. Kunszt and F. Zwirner, \sl Nucl. Phys. \bf B385, \rm 3
             (1992).}
\REF\Gun{J. F. Gunion and L. H. Orr, \sl Phys. Rev. \bf D46, \rm 2052
                    (1992).}
\REF\Kao{H. Baer, M. Bisset, C. Kao, and X. Tata,
         \sl Phys. Rev. \bf D46, \rm 1067 (1992).}
\REF\Stange{V. Barger, K. Cheung, R. J. N. Phillips, and A. L.
               Stange,
         \sl Phys. Rev. \bf D46, \rm 4914 (1992).}
 \FIG\KZregions{Regions of parameter space in which various signatures
           of the supersymmetric Higgs sector can be discovered at
               the LHC, from ref. \KZ.}

Each Higgs boson can decay to a pair of fermions or bosons of any
lighter species, with branching ratios roughly scaling as the
square of the mass of the decay product. Branching ratios to pairs
of supersymmetric particles may be larger if these decays
are kinematically
allowed; in the following discussion, I ignore this possibility.
  In addition, a Higgs boson  can decay to
photon or gluon
pairs through a one-loop amplitude.  These various decay
channels offer many possible signatures that can be observed at a
hadron collider.

 Several
groups have studied the visibility of the variety of Higgs boson
signatures at the LHC\refmark{\KZ--\Stange}.   The dominant decay
to $b\bar b$ is expected to be swamped by hadronic production of
$b\bar b$ pairs.  The most characteristic signature
 of a light Higgs boson
is expected to be the two-photon decay $h^0\rarr \gamma \gamma$.
This decay
is rare and requires high luminosity for its observation, but
it is considered a reasonable target for the LHC experiments.
The decay to $\tau$ pairs can be observed from the sample of
1-prong jets, but the signal is not expected to be sufficient
unless the Higgs coupling to  $\tau^+\tau^-$ is enhanced; this
occurs for $H^0$ and $A^0$ decays to $\tau$ pairs if the parameter
$\tan\beta$ is large.  Finally, if the $H^0$ mass is not too large,
this particle can be observed through its decay to $Z^0 Z^{0*}$,
that is, one on-shell and one off-shell $Z^0$.  In the minimal
supersymmetric standard model, the $H^0$ decouples from $Z^0 Z^0$
if it is as massive as $2m_Z$.   The global picture of the
observability of Higgs decays modes in the minimal supersymmetric
standard model, as a function of the parameters $m_A$ and $\tan\beta$,
 is shown in Fig.  \KZregions.

\REF\chHiggs{M. Felcini, in \sl Proceedings of the
Large Hadron Collider Workshop, \rm  vol. II,
         G. Jarlskog and D. Rein, eds.  (CERN, 1990).}
  Several features of this diagram are worth particular attention.
When this diagram was first presented, it was considered remarkable that,
in most of the plane, there would be some Higgs signal in this
extended Higgs model.  On the other hand, for most choices of parameters,
there is only a single visible Higgs signature, and not necessarily one
that would distinguish this case from the minimal Higgs model.
One would directly observe the heavier Higgs boson $H^0$ and $A^0$
only in certain regions of parameter space; in fact, one is sensitive
to large masses only where the $\tau^+\tau^-$ mode is available at
large $\tan\beta$.  The only known strategy for observing the $H^+$
looks for this particle as a decay product of the top quark and thus
is insensitive to charged Higgs particles with mass above about
120 GeV unless $\tan\beta$ is very large\refmark\chHiggs.

\REF\DaiG{J. Dai, J. F. Gunion, and R. Vega, Phys. Lett. \bf B315,
               \rm 355 (1993), UCD-94-7 (1994).}
\REF\Mrenn{S. Mrenna and G. L. Kane, CALT-68-1938 (1994).}

   Recently, it has been suggested that Higgs decays to $b\bar b$
and $\tau^+\tau^-$ might be made visible at hadron colliders by the
use of multiple vertex tags.\refmark{DaiG,Mrenn}  It remains to be seen
whether this strategy is can be used effectively at high $pp$ luminosity.

  \FIG\Hdecay{The dominant loop diagrams contributing to the decay
           processes $h^0 \rarr gg$ and $h^0\rarr \gamma\gamma$, in
                 the minimal standard model.  In more complex
            models, any additional heavy species can also contribute
           to these amplitudes.}

\REF\GHgam{J. F. Gunion and H. E. Haber, \sl Phys. Rev. \bf D48,
                    \rm 5109 (1993).}
For $m_A$ above 200 GeV, the most important signature is the decay of
the light Higgs boson $h^0$ to two photons.  In principle, the rate
for producing
this signature contains information on the nature of the Higgs boson.
However, that
 information is complicated to extract, since this rate
is proportional to the combination
$$       \Gamma(h^0\rarr gg) \cdot \BR(h^0\rarr \gamma\gamma)\ .
\eqn\ggcombo$$
Both  processes are controlled by loop diagrams, as shown in Fig.
\Hdecay.  Gunion and Haber have argued that the process $h^0\rarr
\gamma\gamma$ is particularly interesting as a probe of exotic
particles\refmark\GHgam.
One might {\it
 assume} that the partial width to $gg$ is dominated by a
top quark loop with standard couplings in order to extract the $\gamma
\gamma$ branching ratio, but it is not clear how to perform an analysis
without  such unwanted assumptions.

  In $\ee$ colliders, the possibilities for observing the Higgs sector
of supersymmetry are much more favorable.  The most important processes
for the production of Higgs bosons are
$$   \ee \rarr   Z^0 \H^0 \ , \qquad      \ee \rarr  A^0 \H^0  \ ,
\eqn\hrxns$$
where $\H^0$ is $h^0$ or $H^0$.
These reactions are complementary in a way that eq.  \Higgsbuild\
makes clear:  A virtual $Z^0$ links a $CP$-even state in the top
line of \Higgsbuild\ to a $CP$-odd state in the bottom line (where
we consider $\pi^0$ to be the longitudinal component of a final state
 $Z^0$).   Whatever linear combination of $h^0$ and $H^0$ is produced
together with $Z^0$, the opposite linear combination is produced
together with $A^0$ when the $\ee$ center of mass energy is
sufficient.  In the minimal supersymmetric standard model, it is
$h^0$ which dominantly couples to
$Z^0$ unless $m_A$ is as small as $m_Z$.

\REF\OritoK{K. Kawagoe and S.  Orito, in \sl Proceedings of the
  Third Workshop on the Japan Linear Collider, \rm A. Miyamoto, ed.
                 (KEK, 1992).}
\REF\JanotH{P. Janot, in \sl
                          Physics and Experiments with Linear
         $\ee$ Colliders, \rm vol. I, F. A. Harris, S. L. Olsen, S.
          Pakvasa, and X. Tata, eds.  (World Scientific, Singapore,
                  1993).}
\FIG\JanotTau{Reconstructed masses of Higgs and $Z^0$ bosons
          from the processes \hrxns.  These simulations, from
         ref. \JanotH, assume 10 fb$^{-1}$ of data at \roots \ =
         400 GeV, with reconstruction efficiencies
           modeled by a LEP scale detector. The shaded region is
       the background, which comes dominantly from $\ee\rarr
                      Z^0 Z^0$. The two figures correspond to
               (a) $m_A = 120$ GeV, (b) $m_A = 180$ GeV.}
  In the $\ee$ environment, it is expected that Higgs bosons, and also
the final state $Z^0$, can be observed in their hadronic  decays modes.
For a collider operating at 400 GeV in the center of mass, the
process $\ee \rarr Z^0 h^0$ is above threshold for any model
arising from grand unification.  This process has a substantial
rate, of order tenths of a unit of R, and is readily
reconstructed\refmark{\OritoK,\JanotH}.
A 500 GeV collider would observe all of the particles in the
supersymmetric Higgs sector for $m_A < 200$ GeV; for higher values of
$m_A$, one need only increase the center of mass energy proportionally.
The analysis which reveals this spectrum can be quite straightforward:
Since the Higgs  bosons and the $Z^0$ all
have $b\bar b$ and $\tau^+\tau^-$
as major decay modes, Janot has suggested looking for
$\ee\rarr (\tau^+\tau^-)(jet\ jet)$ with displaced vertices.  A
simulation based on this search strategy is shown in Fig. \JanotTau\
for two choices the of supersymmetry parameters; the mass peaks of
$h^0$, $H^0$, and $A^0$ are clearly visible.

\REF\Hildreth{M. D. Hildreth, T. L.  Barklow, and D. L. Burke,
           \sl Phys. Rev. \bf D49, \rm 3441 (1994).}
\FIG\Hildrethfig{Expectations for
 Higgs boson branching ratio measurements at a 500 GeV $\ee$ collider,
         from ref.~\Hildreth.  The two figures show simulation
           results for a Higgs boson of mass 140 GeV and 120 GeV,
            assuming 50 fb${^-1}$ of data.  The figures (a) and (b)
             show the theoretical dependence of the branching ratios
              on $m_h$ and $\tan\beta$.}

  Since the $h^0$ is produced with a readily identified $Z^0$, it
should be straightforward to measure the branching fractions to
its major decay modes.  Hildreth, Barklow, and Burke have recently
analyzed this question with simulation studies, assuming a vertex
detector with the capabilities of the one currently operating in the
SLD detector\refmark{\Hildreth}.
 These authors have presented strategies to
isolate the $b\bar b$, $\tau^+\tau^-$, and $W W^*$ decay modes.
A vertex detector closer to the interaction point could also
separate the $gg$ and $c\bar c$ modes, which are predicted to have
have roughly comparable rates.   Janot has proposed a set of cuts
to measure also the branching ratio into invisible final states, for
example, the decay to a pair of neutral supersymmetric
particles\refmark\JanotH.
 The $WW^*$ mode (one real and one
virtual $W$) is particularly interesting because of the relation
between the Higgs production and decay vertices,
$$  { \M(h^0 \rarr  W^+ W^-)\over \M(h^0\rarr Z^0 Z^0)}
     =  \cos^2\theta_w \ ,
\eqn\WWZZrel$$
which assumes only $SU(2)\times U(1)$.  Thus, from the $WW^*$
branching ratio and the total $h^0$ production cross section, one
can compute the $h^0$ total width.  The simulation results of
ref.~\Hildreth\ for the measurement of various branching fractions
are shown in Fig.~\Hildrethfig\ for a data sample
of 50 fb$^{-1}$; branching ratio determinations at $m(h^0)$ = 120 and
140 GeV are plotted against the theoretical dependence on the $h^0$ mass
and $\tan\beta$.  I should note that, in the specific circumstance of
the minimal supersymmetry
standard model, the model-dependence of branching ratios is not as
pronounced as that shown in the bottom figure;
the relative size of the
$b\bar b$ and $W W^*$ branching ratios is proportional to
$$  {\sin^2\alpha \over  \cos^2\beta  \sin^2(\beta-\alpha)}
          =  1 + 2\cos^22\beta \sin^2 2\beta{m_Z^2\over m_A^2} + \cdots
\eqn\ablims$$
in the limit of large $m_A$.

\REF\Telnov{V. Telnov, in
       \sl Physics and Experiments with Linear Colliders, \rm  Vol. II,
       R.~Orava, P. Eerola, and M. Nordberg, eds.
       (World Scientific, Singapore, 1992).}
\REF\BBC{D. L. Borden, D. A. Bauer, and D. O. Caldwell,
        \sl Phys. Rev. \bf D48, \rm4018 (1993).}
     There is one more possible way to study Higgs bosons at
an $\ee$ linear collider:  One can backscatter (visible-light)
laser beams from the electron beams to create a $\gamma\gamma$
collider with approximately the original luminosity and 70-80\%
of the original center-of-mass energy.  In such a facility, the
Higgs bosons can be produced as resonances in $\gamma\gamma$
scattering, and the partial width $\Gamma(h^0\rarr \gamma\gamma)$
can be measured to 10\%\  accuracy\refmark\BBC.  This information
and the $h^0$ total width  would
complement the measurement of the
branching ratio product \ggcombo \ and allows us to determine the
couplings of the $h^0$ both to photons and to gluons.

 For the Higgs sector of supersymmetry, then, the model gives little
impetus to go to
 extremely high energies.  An $\ee$ collider running
at 500 GeV should produce the lightest
Higgs boson in supersymmetric theories---or
in any model where this boson is fundamental at the
grand unification scale---and provide a setting for the detailed
study of its properties.   Such a collider is also, at this
same energy, more likely than the LHC to discover the heavy states of
the Higgs boson spectrum.

\newpart{Supersymmetry: Superpartners}

  In addition to providing an interesting Higgs sector, supersymmetric
models of electroweak symmetry breaking make the characteristic
prediction that the spectrum of elementary particles doubles, with a
new scalar for each species of quark and lepton and a new fermion for
each gauge boson.  If Nature has chosen supersymmetry, the discovery
and characterization of these particles will be the major task of the
next generation of colliders.  In this section, I will compare the
expectations for the ability of $\ee$ and $pp$ colliders to investigate
these new particles.

   To introduce this comparison, I will review some general
features of
the expected mass spectrum  and decay patterns of supersymmetric
particles.  The phenomenology of supersymmetry is often
discussed in a framework in which the model is viewed as part of a
grand unified theory with the simplest pattern of supersymmetry
breaking.  In my discussion, I will use these assumptions to
make rough estimates of the mass relations among supersymmetric
particles.  I will ask whether deviations from these assumptions,
which are after all very likely, are observable experimentally.
Since supersymmetry is a weak-coupling theory, one can derive
detailed predictions from simple assumptions, and it is seductive
to consider these predictions as resting on a firm footing.
Some of the predictions are, in fact, quite robust with respect
to changes in the assumptions; I will point out examples below.
Other predictions can change dramatically.  From the viewpoint of
recommending future colliders, these latter predictions have special
interest, because they lead to experimental probes of the
mechanism of supersymmetry breaking which comes down from the
unification or gravitational scale.

   In the simplest type of supersymmetric grand unification,
the masses of superpartners are controlled by three mass parameters:
$m_0$, a universal scalar mass, $m_{1/2}$, a universal gaugino
mass, and $\mu$, a supersymmetric Higgs boson mass parameter.
All three masses are roughly of the size of $m_W$; a reasonable
theory of supersymmetry breaking should explain their near equality.
The universality of $m_0$ and $m_{1/2}$ refers to their values at the
scale of grand unification; at lower energies, the masses of
different species may differ due to renormalization.  For example,
in the approximation of one-loop renormalization group equations,
the masses of the fermionic partners of the  gauge bosons
of $SU(3)\times SU(2)\times U(1)$ are expected to obey
$$  {m_1\over \alpha_1} = {m_2\over \alpha_2 } = {m_3\over \alpha_3}
=  {m_{1/2}\over \alpha_{\rm GUT}} \ ,
\eqn\mhalfrels$$
 where  the coupling constants are as in grand unification:
$$\alpha_3 = \alpha_s\ ,\qquad \alpha_2 = {\alpha\over\sstw} \ ,
\qquad   \alpha_1  = {5\over 3} {\alpha\over \cstw}  \ .
\eqn\whatarealphas$$
One consequence of this renormalization is that the (mass)$^2$ of the
Higgs field $\phi_2$, which begins at $m_0^2$  at the grand unification
scale, becomes negative at low energy due to loop diagrams involving
the top squarks.  This is in fact the mechanism
of $SU(2)\times U(1)$ breaking in supersymmetric models.

\REF\Barbieri{R. Barbieri, in \sl Z Physics at LEP 1, \rm G. Altarelli,
         R. Kleiss, and C. Verzegnassi, eds, vol. 2.
          (CERN, Geneva, 1989).}
\REF\Ross{G. G. Ross and R. G. Roberts, \sl Nucl. Phys. \bf B377,
        \rm 571 (1992).}
\REF\ArnN{R. Arnowitt and P. Nath, \sl Phys. Rev. Lett. \bf 69,
               \rm 725 (1992).}
\REF\TAM{S. Kelley, J. L. Lopez, D. V. Nanopoulos, H. Pois, and
           K. Yuan, \sl Nucl. Phys. \bf B398, \rm 3 (1993).}
\REF\Kane{G. L. Kane, C. Kolda, L. Roszkowski, and J. D. Wells,
       UM-TH-93-24 (1993).}

  These general ideas lead to a qualitative picture of the superparticle
mass spectrum.  Most importantly, $m_W$ is a scale of supersymmetry.
This scale is generated by the same mass terms which give mass to the
superpartners.  While it is possible to adjust the parameters of
the theory so that $m_W$ is light while the underlying mass
parameters  are much heavier, this situation is unnatural.   In
specific models which incorporate this physics, the lighter
supersymmetric
partners of the $W$ and $Z$ typically have masses below about 200 GeV,
with other superpartner masses scaling
accordingly\refmark{\Barbieri--\Kane}.

  The second aspect of this picture is that color singlet superpartners
are typically much lighter than colored superpartners. From eq.
\mhalfrels, we see  that the gluino, the partner of the gluon,
is by far the heaviest gauge fermion.   I will give a precise
statement of this relation below.
The relation between squark and slepton masses is more model-dependent.
At the level of one-loop renormalization group equations, and assuming
that both squark and gluino masses are much larger than $m_Z$,
 the
squark masses at the weak scale obey:
$$         m^2(\widetilde q) \simeq  (0.7 m_3)^2 + m_0^2  \ .
\eqn\squmasses$$
The first term arises from the squark mass renormalization due to gluino
loops.  Sleptons
acquire a similar, but smaller, mass correction from loop diagrams
involving
the weak gauge fermions.  If the $m_0$ term dominates eq. \squmasses,
then both squarks and sleptons will be very heavy; in the opposite
limit, the slepton masses will be of the same order as the masses
of the $W$ superpartners.
These renormalizations also lead to mass splittings between the
squarks and sleptons associated with right- and left-handed fermions,
even for the case of universal scalar masses at the unification scale.
These splitting are of order 5\%
for squarks but should be large for
light sleptons.  For example, this model predicts the relation
$$    m^2(\widetilde\ell_L) - m^2(\widetilde \ell_R) =
   (0.6 m_{1/2})^2 \ ,
\eqn\melleq$$
up to negligible terms proportional to $(1-4\sstw)m_Z^2$.

If we relax the assumption that
the scalar masses are  universal, many of these detailed results
can be upset.
 Some of the predictions do remain valid in a more
general context; in particular, the large positive mass shifts for the
gluino and the squarks follow from the renormalization group
equations almost irrespective of their initial conditions.
On the other hand, the near degeneracy of the squark masses, and the
specific pattern of the slepton masses, depends crucially on the model
assumptions.  If Nature has chosen supersymmetry, we must be able to
test these assumptions, and we certainly cannot rely upon them.

   \REF\options{For discussion of other options, see ref. \HandK\ and
          S. Dimopoulos and L. Hall, \sl Phys. Lett. \bf 207B, \rm
                   210 (1988).}
 The characteristic signatures of supersymmetry at colliders
involve the decay of heavier supersymmetric species into the lightest
superpartners.  Thus, I should begin a discussion of these signatures
by reviewing the properties of these lightest states.  There are
several possibilities for the lightest superparticle: This particle
might be a neutral fermion or the scalar partner of the neutrino,
and it may or may not be stable with respect to decay to more
familiar particles.  In this discussion, I will make a particular
choice---the one which is least problematical and most thoroughly
analyzed---that
the lightest superpartner (LSP) is a
neutral fermion, and that it is absolutely stable\refmark\options.

  Under this hypothesis, the LSP is a linear combination of the
fermionic partners of the photon, the $Z^0$, and the two neutral
Higgs fields $\phi_1^0$ and $\phi_2^0$.  Supersymmetry requires
that these four states mix
with one another in a complex pattern.  The mass eigenstates
are given by diagonalizing the following matrix, written in the
$SU(2)\times U(1)$
basis  $(\widetilde B, \widetilde A^3, \widetilde \phi_1^0,
\widetilde \phi_2^0)$:
$$  \pmatrix{m_1 & 0 & -m_Z\sin\theta_w\cos\beta&
             m_Z\sin\theta_w \sin\beta \cr
             0 & m_2 &
m_Z\cos\theta_w\cos\beta&
          -   m_Z\cos\theta_w \sin\beta \cr
 -m_Z\sin\theta_w\cos\beta&
             m_Z\cos\theta_w \cos\beta &
             0 &- \mu \cr
m_Z\sin\theta_w\sin\beta&
          -   m_Z\cos\theta_w \sin\beta &
             - \mu & 0 \cr } \ .
\eqn\neutralino$$
The parameters $m_1$, $m_2$, and $\mu$ are determined (or not, as
Nature chooses) by the unification relations described above.
The entries which are proportional to $m_Z$ are determined by the
 supersymmetry relations between the couplings of the Higgs fields
and those of their fermionic partners.   The
eigenvectors of this matrix correspond to four
massive fermions which are
called {\it neutralinos}.  The heavier neutralinos typically decay
to the lightest one, the LSP, by emitting weak bosons or quark
or lepton pairs.

\FIG\CharginoPlane{Contours of constant mass of the lighter
        chargino $\widetilde\chi_1^+$ in the $(\mu, m_2)$ plane.
         The two curves correspond to
          $m(\widetilde\chi_1^+) =$ 125, 250 GeV, and $\tan\beta = 4$.
        The labels indicate the regions in which the lightest
         charginos and neutralinos are mostly gauge boson or mostly
            Higgs boson superpartners.}

   Similarly, the fermionic partners of the $W$ and charged Higgs
bosons mix.  This mixing problem is best described by considering
partners of $W^+$ and $\phi_2^+$ as the left-handed components of
Dirac fermions, while the antiparticles of the partners of
$W^-$ and $\phi_1^-$ are the corresponding right-handed components.
This leads to a mass matrix
$$   \pmatrix{\widetilde w^- & \widetilde \phi_1^-\cr}
    \pmatrix{m_2 & \sqrt{2} m_W \sin\beta \cr
    \sqrt{2} m_W \cos\beta &  \mu\cr}
   \pmatrix{\widetilde w^+ \cr \widetilde \phi_2^+\cr} \ .
\eqn\chargino$$
The mass eigenstates are called {\it charginos}.

    \REF\Pierce{S. P. Martin and M. T. Vaughn, \sl Phys. Lett.
           \bf B318, \rm 331 (1993); D. Pierce and A.
            Papadopoulos, JHU-TIPAC-940001 (1994).}
  The lighter of the two
charginos, $\widetilde \chi^+_1$, has a (mass)$^2$ less than
$(m_2^2 + m_W^2)$.    The
unification relation between $m_2$ and $m_3$,  eq. \mhalfrels,
relates this bound to the mass of the
gluino.  Taking account of the fact that the physical or `pole'
mass of the gluino is  15-20\% higher than the mass $m_3$ due to
QCD corrections\refmark{\Pierce},  we find
$$ m^2(\widetilde \chi_1^+) < \bigl({1\over 4}  m(\widetilde g)\bigr)^2
           +  m_W^2 \ .
\eqn\widemasses$$
This relation and eq. \squmasses\
quantify the remark made earlier that the color singlet
superparticles are typically much lighter than the colored
superparticles.

The properties of the two chargino eigenstates  depend on the
relative sizes of all of the parameters in \chargino.  In  Fig.
\CharginoPlane, I display  contours of constant mass for
$\widetilde \chi^+_1$.  In the regions indicated, the lightest
chargino and neutralino are
mainly gaugino or mainly Higgsino.  From one region
to another, the decay properties of the chargino and of heavier
superparticles change qualitatively.  We will
see the consequences of this in a  moment.

 Though the mixing problem of the charginos and neutralinos is complex,
there are a few simple features which emerge.  Eventually, the
heavier charginos and neutralinos, and other heavy superparticles,
will decay down to the LSP.  This particle then escapes detection,
leading to  missing  transverse momentum and energy.
If superpartners have a large production cross section, this signature
is robust across the parameter space and is readily observed.

 \REF\SDCTC{Solenoidal Detector Collaboration (SDC), \sl Technical
               Design Report. \rm (SSC Laboratory, 1992).}
\REF\LHCA{G. Polesello, in \sl International Workshop on Supersymmetry
        and the Unification of the Fundamental Interactions,  \rm
            P. Nath, ed.  (World Scientific, Singapore, 1993).}
  \FIG\FromSDC{Spectrum of missing transverse energy expected in
          the SDC detector at the SSC, due to production of gluinos
                with mass 300 GeV and 500 GeV, from ref. \SDCTC.}
  One can search for this missing energy signature quite
  straightforwardly at high energy $pp$ colliders. Since the gluino is
a color octet fermion, it has a large production cross section
in $gg$ collisions.  The decay products of the gluino include
the LSP, which gives rise to events with missing energy.
The spectrum of observed  missing transverse energy, together with
a background estimate made for the SDC detector at the SSC, is shown
in Fig. \FromSDC.  The ATLAS collaboration has estimated that this
signature is visible at the LHC up to gluino masses of about 1.6 TeV,
assuming a data sample of 100 fb$^{-1}$,
even if squarks are much heavier than gluinos\refmark\LHCA.
This goes about a factor 2 beyond the rough theoretical upper limits
discussed above.

\REF\BTW{H. Baer, V. Barger, D. Karatas, and X. Tata,
             \sl Phys. Rev. \bf D36, \rm 96 (1987).}
\REF\BGH{R. M. Barnett, J. F. Gunion, and H. E. Haber,
             \sl Phys. Rev. \bf D37, \rm 1892 (1988).}
\REF\tat{H. Baer, X. Tata, and J. Woodside,
             \sl Phys. Rev. \bf D42, \rm 1568 (1990).}
\REF\tattwo{H. Baer, X. Tata, and J. Woodside,
             \sl Phys. Rev. \bf D45, \rm 142 (1992).}
\FIG\BGHchange{Variation of the gluino branching fractions as a
           function of $\mu$, for fixed $m_2$, $m_3$,
            from ref. \BGH.  The three curves show the branching
            fractions for direct decay  to the LSP, decays to on-shell
            Higgs bosons, decays to on-shell $W$ and $Z$ bosons,
             and decays to 5-particle final states.  The structure
        in the center of each diagram is the result of transitions
         between the gaugino region and the Higgsino region of
                Fig. \CharginoPlane.}
  Once the gluino signature is found, can one establish that this new
particle is a superpartner and use its decays to study supersymmetry.
It is true that the gluino is free of the mixing problems that we
found with the neutralinos, and that it is its own antiparticle.
But all other features of the gluino decay are exceedingly complicated.
I have already pointed out that the gluino is expected to be much
heavier
 than the lightest neutralinos and charginos.  Thus, the
gluino is expected to decay not only to the lightest particle in this
sector but also to the heavier gauge partners, which then decay to
the LSP through a complicated decay chain. For example, a decay through
the next heaviest superpartners leads to  the processes such as
$$ \eqalign{
        \widetilde g \rarr &  q \bar q \widetilde\chi^+_1 \cr
       \rarr &  q \bar q   \quad q \bar q \widetilde\chi^0_1\ . \cr}
\eqn\cascade$$
 In the these decays and the
similar decays to neutralinos, the intermediate steps involve
virtual (or real) $W$,
 $Z$, or Higgs bosons, or virtual squarks and sleptons.
   In simulations of these
decay chains, the direct decay to the LSP turns out to be rare,
while decays through two or more intermediate superpartners are
quite common\refmark{\BTW--\tattwo}.   The model-dependence of the
branching fractions of the gluino into various final states is
illustrated in Fig. \BGHchange.

\FIG\Tatas{Cross sections expected at the LHC for a variety of
         signatures of gluino production, from ref. \tattwo.
             The various curves show the cross sections for
             missing transverse energy, same sign dileptons, and
             production of the indicated numbers of on-shell $Z$
                bosons and isolated leptons, (a) as a function of
                the gluino mass for $\mu$ fixed at $-150$ GeV, (b)
                 as a function of $\mu$ for a gluino mass fixed
                   at 750 GeV.}
  The complexity of gluino decays has advantages and disadvantages.
On the one hand, it leads to a wide variety of gluino signatures,
including multilepton  and lepton + $Z^0$ signals in addition to
missing $E_T$.  The expected cross sections for
 these signatures at the LHC,
computed at a particular point in the parameter space of the
neutralinos and charginos, is shown in Fig. \Tatas.  Along with this
feature, one cannot avoid the difficulty that  the strengths of these
signals depend on the properties of the charginos and neutralinos and
therefore on the full complexity of the mixing problems
\neutralino\ and \chargino.  The lower graph in Fig. \Tatas\ shows
the dependence of the signatures on $\mu$ for fixed gluino mass.

\REF\BGHm{R. M. Barnett, J. F. Gunion, and H. E. Haber, in
        \sl Research Directions for the Decade
           (Snowmass, 1990), \rm E. L. Berger, ed.
              (World Scientific, Singapore, 1992).}
\FIG\BGHmass{A jet mass combination which estimates the gluino mass,
         from ref.~\BGHm.  In a simulation for the SSC, events with
            two isolated like-sign leptons are selected.  Then the
       momentum vector of the highest-$p_T$ lepton is combined with
        those of the two nearest jets chosen from the four highest-$p_T$
         jets.  The resulting mass distribution is shown for a gluino
          of mass 300 GeV and 350 GeV.  (The latter histograph is shown
          divided
            by two).}
\REF\Tataandco{H. Baer, M. Drees, C. Kao, M. Nojiri,   and
X. Tata, FSU-HEP-940311 (1994).}
 In favorable circumstances, some features of these events can provide
important pieces of information.  Barnett, Gunion, and Haber have pointed
out that by combining the momentum vectors of the highest $p_T$
lepton and the two closest jets, one obtains an estimator for the
gluino mass.  The mass resolution is expected to be about 10\%, as shown
in Fig. \BGHmass.  Since the gluino is its own antiparticle, the two
hardest leptons are expected to be  of like or unlike
sign with equal probability,
 and this property distinguishes  supersymmetry from
other possible models of new colored  fermions.  However, neither
experiment is unambiguous.   Even discounting standard model
backgrounds and misidentifications, supersymmetry itself offers many
other sources of leptons, for example from squark decays or from the
lower stages of $\widetilde \chi$ cascades.  Under specific
 circumstances, such as the presence of a light top squark, these
new sources not only confuse but actually swamp the more direct
lepton signals\refmark\Tataandco.

\REF\Tchargino{J. L. Lopez, D. V. Nanopoulos, X. Wang, and A.
             Zichichi, \sl Phys. Rev. \bf D48, \rm 2062 (1993).}
\REF\Hchargino{H. Baer, C. Kao, and X. Tata,
        \sl Phys. Rev. \bf D48, \rm 5175 (1993); H. Baer,
                C. Chen, F. Paige, and X. Tata, FSU-HEP-940310 (1994).}
  Searches for other superparticles have also been considered at
hadron colliders, and these have many of the same opportunities and
the same problems that we have seen for the gluino.  The production
cross section for squarks is similar to that for the gluino, and the
technique for mass measurement is similarly indirect.  In fact, there
is no published method for distinguishing the cases in which squark
or gluino production is dominant (though this is certainly a solvable
problem).  Recently, several groups have studied chargino searches
in $q\bar q$ annihilation at hadron colliders; this is an interesting
production method for relatively light charginos, though it
disappears behind the background for chargino masses
 above about 150 GeV\refmark{\Tchargino,\Hchargino}.

\REF\Grivaz{J.-F. Grivaz,   in
       \sl Physics and Experiments with Linear Colliders, \rm  Vol. I,
       R.~Orava, P. Eerola, and M. Nordberg, eds.
       (World Scientific, Singapore, 1992).}

 We now turn to supersymmetry signatures at $\ee$ colliders.  In
principle, any superpartner with nonzero electroweak quantum numbers
is produced in $\ee$ annihilation with a substantial cross section.
For the purpose of comparison with $pp$ colliders, it is important to
note that $\ee$ colliders have no difficulty in producing the
color singlet superparticles such as charginos.  The chargino signal
is simple and easily isolated, as has been
discussed, for example, by
Grivaz\refmark\Grivaz.
  Because these
particles are expected to be lighter than their colored counterparts,
 according to the relation
\widemasses, an $\ee$ collider even at 500 GeV in the center of mass
is sensitive to a region of supersymmetry parameter space similar
to that of a  search up to a gluino masses of 1 TeV.  This already
covers the region expected from the theoretical considerations described
above.    If one
is concerned to push beyond this region, an
$\ee$ collider at 1 TeV is actually sensitive to a larger region of
supersymmetry parameter space than the LHC.

 \REF\OritoH{S. Orito, in
                         \sl  Physics and Experiments with Linear
         $\ee$ Colliders, \rm vol. I, F. A. Harris, S. L. Olsen, S.
          Pakvasa, and X. Tata, eds.  (World Scientific, Singapore,
                  1993).}
\REF\TFMYO{T. Tsukamoto, K. Fujii, H. Murayama, M. Yamaguchi,
and Y. Okada, KEK-PREPRINT-93-146 (1993).}

 However, the most important advantages of an $\ee$ collider become
apparent at the next stage, when one has found the first signal of
supersymmetry.  We have seen above that it is very difficult to
translate the signatures seen in
$pp$ collisions to definite knowledge
of the supersymmetry parameters.   In $\ee$ collisions, the situation
is quite the reverse: One can build up knowledge of the supersymmetry
parameters systematically and straightforwardly\refmark\OritoH.
  There are two
important reasons for this special simplicity.  First, an $\ee$
collider can directly produce the lightest states of the superparticle
spectrum and then characterize the spectrum in stages of increasing
mass.  Second, an $\ee$ collider offers incisive probes of the
superspectrum not available at hadron colliders, especially the
handle of electron beam polarization.  Tsukamoto, Fujii,
Murayama, Yamaguchi, and Okada have clarified this strategy by
presenting simulation results on the determination of supersymmetry
parameters for a particular choice of the superpartner
masses\refmark\TFMYO;  my discussion will draw
strongly on their work.

\FIG\CHmass{Energy distribution of the 2-jet system produced in the
       decay  $\widetilde\chi^+ \rarr \widetilde\chi^0 q \bar q$,
          as reconstructed in the simulations of ref. \TFMYO.  This
          work assumes an integrated luminosity of 20 fb${}^{-1}$
        at an $\ee$ linear collider operating at \roots\ = 500 GeV.
         The right-hand figure shows the $\chi^2$ distribution for the
         reconstructed masses of $\widetilde\chi^+$ and
           $\widetilde\chi^0$.}
  In contrast to the complex decay pattern we have seen for gluinos,
the lightest states in the superspectrum have only a single allowed
mode of decay to the LSP.  The production mechanism is also much
simpler, since the new particles are pair-produced and thus have an
energy which is precisely defined, up to minor effects of initial-state
radiation.  Then  the masses of the new state and that of the LSP can
be deduced from the endpoints of the energy distribution of
observed products.  The simplest example is given by the
superpartner of the $\mu_R^-$,
 for which the
major decay mode is the 2-body decay $\widetilde \mu_R \rarr \mu +
\widetilde\chi^0_1$.  The muon energy distribution is flat between the
endpoints, which can then be read off to an accuracy of 1 GeV.  A
more typical example is that of the lightest chargino.  Fig. \CHmass\
shows the reconstruction of the  dijet energy distribution in the
decay $\widetilde\chi^+_1 \rarr q \bar q \widetilde \chi^0_1$,
assuming the detector model of the JLC group.  The masses of the
$\widetilde\chi^+_1$ and the $\widetilde\chi^0_1$  are each
determined to an accuracy of 2 GeV.

\FIG\Polarize{Feynman diagrams contributing to the
production of selectron and chargino pairs at $\ee$ colliders.
These processes illustrate the simplifications obtained by
controlling the electron beam polarization.}

Once we have determined the masses of the lightest charginos and
neutralinos, we will also need to determine the mixing angles which
relate the underlying basis of superpartners to the mass eigenstates.
To some extent, these can be determined from production angular
distributions, but the use of electron beams with definite
polarization can provide wonderful simplifications.  Two reactions
which illustrate these simplifications are shown in Fig. \Polarize.
In selectron production, shown in Fig. \Polarize(a), the $t$-channel
diagram exists only if the final selectron is the superpartner of the
initial electron; thus, choosing $e_R^-$ selects $\tilde e_R^-$.
Since the $e_R^-$ is a singlet of weak interaction $SU(2)$, the
first diagram involves only the $U(1)$ gauge boson $B^0$,
 a linear combination
of $\gamma$ and $Z^0$.  Similarly, the $t$-channel diagram involves only
the superpartner $\widetilde B^0$ of this boson.  By measuring the
contribution to the $t$-channel amplitude from each massive neutralino,
one measures the mixing angle between each mass eigenstate
and the $\widetilde B^0$.

\REF\OurCh{J. Feng, H. Murayama, M. E. Peskin, and X.
Tata, in preparation.}
  Similarly, in chargino production,
Fig. \Polarize(b), the choice of $e_R^-$ removes the $t$-channel
diagram involving sneutrino   exchange and allows the measurement of
both mixing angles of the chargino.  To make this plausible, I will
quote the formula for the angular distribution of chargino pair
production
 in the asymptotic limit
limit $\sqrt{s} \gg
m_Z, m(\widetilde\chi^+_1)$:
$$  {d\sigma\over d\cos\theta}
(e_R^- e^+\rarr \widetilde\chi^+_1 \widetilde\chi^-_1)
\sim   \sin^4 \phi_+  (1+\cos\theta)^2
         + \sin^4\phi_- (1 - \cos\theta)^2 \ ,
\eqn\chiangdist$$
where
 $\phi_+$ and $\phi_-$ are the mixing angles relating the
basis of electroweak and mass eigenstates on the right- and left-hand
sides of \chargino.
When the second chargino is discovered, in the
reaction
$\ee\rarr \widetilde\chi^+_1 \widetilde\chi^-_2$, the masses of the
two charginos $M_1$, $M_2$
can be combined with the two mixing angles to reconstruct
the off-diagonal elements of the chargino mass matrix.  To first
order in the mixing angles, this relation reads
$$  2 m_W^2 =  (M_1^2 + M_2^2)(\phi_+^2 + \phi_-^2) - 4 M_1 M_2
                         \phi_+\phi_-  \ .
\eqn\masstest$$
We noted below \neutralino \ that $m_W$ appears in the chargino
mass matrix by virtue of a supersymmetric relation of couplings.
Thus, eq. \masstest\ provides
a simple quantitative test of
supersymmetry which can be realized in a large part of parameter
space.  In other regions, one can alternatively test the
supersymmetry relation for the electron-selectron-neutralino and
electron-sneutrino-chargino couplings\refmark\OurCh.

\FIG\Uniftest{Supersymmetric grand unification tests available
       at a 500 GeV linear collider in the scenario studied in
ref. \TFMYO: (a) the test of the  mass
relation  \mhalfrels\
 between the $SU(2)$ and $U(1)$ gauge boson superpartners;
(b) the test of the  mass relation \melleq\
involving the $SU(2)$ gauge boson and the electron superpartners.
Both relations require the determination of mixing angles as well
as physical particle masses. The two
 $\chi^2 = 1$ contours include consideration of
 reconstruction efficiencies and standard model
 backgrounds.}

\REF\FengandF{J. L. Feng and D. E. Finnell,
        \sl Phys. Rev. \bf D49, \rm 2369 (1994).}
  Another use of the determination of chargino and neutralino mixing
angles is to test the unification assumptions discussed at the
beginning of this section.  Tsukamoto, \etal \ studied the extent to
which one could use the values of the mixing angles obtained from
their simulation results to test the relation \mhalfrels\ between
$m_1$ and $m_2$ and the relation  \melleq\ relating the masses of the
partners of $e_L^-$ and $e_R^-$.  Their results are shown in Fig.
\Uniftest.  Feng and Finnell have shown that it is also
possible to
measure mass differences between left- and right-handed quark
superpartners at an $\ee$ collider at the level of a few percent,
by making use of the
polarization-dependence of cross sections\refmark\FengandF.

  These tests of supersymmetry or supersymmetric unification
have no analogue completely within the domain of experiments at $pp$
colliders.  However, it is exceedingly interesting to test the
relation between $m_2$ and $m_3$, and relations between squark and
slepton masses, bringing together information from the two types
of facilities.  Even more importantly, the measurement of
chargino and neutralino masses and mixing angles at $\ee$ colliders
will provide an experimental basis for the modeling of squark and
gluino cascade decays at hadron colliders.  This step may be
essential to the process of converting data on hadronic signatures of
supersymmetry to quantitative knowledge of the supersymmetry
parameters.

\newpart{Technicolor}

  Now that we have reviewed the experimental prospects for supersymmetry
models in some detail, let us turn to the experimental study of
technicolor models.  Technicolor provides a  concrete setting in which
strong-coupling physics leads to electroweak symmetry breaking.
It is a complete theory, and so it also leads to a variety of
observable phenomena which can shed light on the
various aspects of the new sector of physics.  In this section, I will
compare $\ee$ and $pp$ colliders in their ability to observe these
various manifestations of technicolor.    The most important questions
to be answered experimentally are summarized in Table 2.

In the case of supersymmetry, we were able to discuss detailed and
quantitative predictions for many phenomena.  In the case of
technicolor, the predictions we discuss will be more limited.
There are two reasons for this:
 First, the theory includes strong-coupling
phenomena, and thus
theorists are limited at present to  semiquantitative
means of calculation.
Second, there is no `technicolor standard model' which is simple,
compact, and consistent with all present data.  The simplest
technicolor models have serious phenomenological problems, and
{\it ad hoc} cures for these problems may remove or alter some of the
generic signatures of technicolor.

  Since technicolor is based on new strong interactions, the most
striking signature of technicolor would be the discovery of
the resonances of those interactions.  In technicolor models, these
resonances are assumed to mirror those of the familiar strong
interactions.  The states analogous to
 the pions are eaten by the $W$ and $Z$   to form their
longitudinal components.  If the flavor symmetry of technicolor is
larger than $SU(2)\times SU(2)$, there will be additional observable
pseudoscalar mesons, which I will discuss below.  But the general,
characteristic signal of technicolor is the appearance of an
$SU(2)$ triplet of rho resonances.  Because the weak bosons contain
a techni-pion component, these resonances should appear in $WW$
and $WZ$ scatting in the channel with isospin and spin
$I = J = 1$.  The mass of the
techni-rho is expected to be between 1 and 2 TeV; the  resonance
is expected to be relatively narrow, with a width of a few
hundred GeV.

 \REF\CG{M. S. Chanowitz and M. K.  Gaillard, \sl Nucl. Phys.
            \bf B21, \rm 379 (1985).}
\REF\Chano{M. S. Chanowitz and W. Kilgore, \sl Phys. Lett. \bf B322,
                 \rm 147 (1994).}
\REF\Hans{J. Bagger, \etal, \sl Phys. Rev. \bf D49, \rm 1246 (1994).}
\REF\AachenR{I. Josa, F. Pauss, and T. Rodrigo, in
             \sl Proceedings of the
Large Hadron Collider Workshop, \rm  vol. II,
         G. Jarlskog and D. Rein, eds.  (CERN, 1990).}
\REF\Orsay{F. Iddir, A. Le Yaouanc, L. Oliver, O. Pene, and J. C.
               Raynal, \sl Phys. Rev. \bf D41, \rm 22 (1990).}
\REF\MyFin{M. E. Peskin, in
  \sl Physics and Experiments with Linear Colliders, \rm Vol. I,
       R.~Orava, P. Eerola, and M. Nordberg, eds.
       (World Scientific, Singapore, 1992).}
\REF\TimBTR{T. L. Barklow, in
       \sl Physics and Experiments with Linear Colliders, \rm Vol. I,
       R.~Orava, P. Eerola, and M. Nordberg, eds.
       (World Scientific, Singapore, 1992).}

\FIG\TRRES{Behavior of the differential cross section for
$\ee\rarr\pairof{W}$,
 $d\sigma/d\cos\theta$ at $\cos\theta = 0$, as a function of
\roots, under in a theory with a techni-$\rho$ resonance at the given
mass, from ref. \MyFin.  The cross section is given in units of R.}

\FIG\TRtim{Projected
sensitivity of measurements of the differential cross  for
    $\ee\rarr W^+W^-$ to the presence of a techni-rho resonance,
           from ref. \TimBTR.
       The figure shows the expected constraints on the technicolor
          analogue of the pion form factor at $Q^2 = s$, and the
         corresponding predictions for various techni-rho masses.
         The two figures correspond to a 1 TeV $\ee$ collider
            with integrated luminosity 200 fb$^{-1}$ and a
               1.5 TeV collider with 500 fb$^{-1}$.}

   The physics studies for the SSC included substantial work on the
 question of experimentally observing
the scattering of weak interaction
bosons\refmark{\CG--\Hans}.  Among the various models considered,
the assumption of a narrow techni-rho provides a relatively
easy target.  In ref. \AachenR, it is shown that
the techni-rho can be reconstructed as a clear
 peak in the $WZ$ invariant
mass spectrum at the LHC (assuming its highest luminosity) if the
techni-rho mass is 1.5 TeV; the signal disappears below the background
for a  techni-rho mass above of 2 TeV.

    \REF\Wcross{W. Beenakker, A. Denner, S. Dittmar, R. Mertig, and
               T. Sack, \sl Nucl. Phys. \bf B410, \rm 245 (1993);
               W. Beenakker and A. Denner,  DESY-94-051 (1994).}
  The techni-rho is also particularly straightforward to observe
in $\ee$ annihilation.  Just as the conventional rho meson is the
dominant effect in the  pion form factor, the techni-rho
creates a dramatic enhancement in the cross section for $\ee\rarr
\pairof{W}$, which is the most important single
process in $\ee$ annihilation at high energy\refmark{\MyColl,\Orsay}.
The effect of a techni-rho resonance on the differential cross section
for $W$ pair production at 90$^\circ$, for techni-rho masses of 1 and
1.5 TeV,
is shown in Fig. \TRRES.  Even if the techni-rho resonance is located
at much higher energy, its effects are observable if one relies on
the ability of $\ee$ colliders to reconstruct $W$ bosons and measure
their polarization, and on the fact that the standard model prediction
for the $W$ pair production cross section is known to better than
1\% accuracy\refmark\Wcross.  The sensitivity of $\ee$ colliders
to techni-rho resonances of very large mass is shown in Fig. \TRtim.

\REF\KEKfolks{Y. Kurihara and R. Najima, \sl Phys. Lett. \bf B301,
                       \rm 292 (1993).}
  I should note that this sensitivity to weak boson resonances is
special to effects in the $I=J=1$ channel.  There is no analogous
reaction which is sensitive to a narrow resonance in the $I=J=0$
channel.  However, there are also no known models which produce
such an effect; for example, a minimal Higgs boson at 1.5 TeV
has a width of order its mass.  In the case of a broad resonance
in this channel,
the weak boson scattering can be observed at an $\ee$ collider
just as at the LHC\refmark{\KEKfolks},  but the experiment is
difficult at both colliders.

\REF\MypGb{M. E. Peskin, \sl Nucl. Phys. \bf B175, \rm 197 (1980).}
\REF\JohnspGb{J. P. Preskill,
          \sl Nucl. Phys. \bf B177, \rm 21 (1981).}
\REF\Holdom{B. Holdom, \sl Phys. Rev. \bf D24, \rm 1441 (1981).}
\REF\Appelwalk{T. Appelquist, D. Karabali, and L. C. R. Wijewardhana,
           \sl Phys. Rev. Lett. \bf 57, \rm 957 (1986).}
\REF\ELnew{K. Lane and M. V. Ramana, \sl Phys. Rev.  \bf D44,
                 \rm 2678 (1991).}
   If there are more than two flavors of techni-fermions, the
technicolor model will produce additional pseudoscalar mesons
which are not eaten by the $W$ and $Z$ and which therefore
appear as physical particles.  These mesons, which
unfortunately have the name {\it pseudo-Goldstone bosons} (pGb's),
look much like  the
$CP$-odd Higgs bosons of extended Higgs sectors.
  They are expected to couple most
strongly to the heaviest flavors---$\tau$, $b$, and $t$.
If technifermions carry the color $SU(3)$ of the familiar strong
interactions, pGb's may carry color and so may be produced in
$pp$ collisions.  They are visible as new sources of high transverse
momentum top quark or $\tau$ lepton pairs,
up to masses  of about 1 TeV at the LHC\refmark{\EHLQ}.
At the same time, pGb's are expected to form singlets and
triplets of weak
$SU(2)$, and the triplets can be discovered at $\ee$ colliders,
in any favored decay mode,
up to the pair-production threshold.

   To evaluate the relative strength of $pp$ and $\ee$ colliders
for this study, it is necessary to have some idea of the masses
expected for pGb's.  For those pGb's which carry
 strong interaction quantum numbers, it is easy to compute that standard
model radiative corrections give them  masses of order 200
GeV\refmark{\MypGb,\JohnspGb}.  However, if only these corrections
are included, technicolor models have the dual problems of
containing very light charged scalar particles and flavor-changing
neutral currents.  The natural solution to these problem, due to
Holdom\refmark{\Holdom,\Appelwalk},
 leads to
an additional contribution to the pGb masses which is difficult
to calculate and which may be as large as the
techni-rho mass.
  However, Lane and Ramana
have recently proposed a variant of this model
which contains both colored and charged pGb's  below 300
GeV\refmark{\ELnew}.

   Once one has invoked new strongly interacting fermions to break
electroweak symmetry, it is still necessary to convey this symmetry
breaking to obtain masses for the quarks and leptons.  In
conventional technicolor models, the bridge between the technifermions
and the ordinary fermions is made by an additional new set of
interactions, called extended technicolor.  These new interactions
naturally live at scales of order 100 TeV, or even higher in models
based on Holdom's ideas.  However, the large mass of the top quark
requires that at least the particular boson responsible for this mass
should have a mass at the 1 TeV scale and also be relatively strongly
coupled.  Thus, this particular ETC boson should be a target of
direct and indirect searches at the next generation of colliders.

\REF\AWETC{P. Arnold and C. Wendt, \sl Phys. Rev. \bf D33, \rm 1873
                  (1986).}

   The question of direct searches for ETC bosons was studied
some time ago by Arnold and Wendt\refmark{\AWETC}.
 The lightest ETC boson carries both top quark and technifermion
quantum numbers.  Thus, it must be pair-produced, and the
production process is dominated by $s$-channel resonances of
ETC and anti-ETC bosons bound by technicolor forces.  This state
then decays to lower mass technicolor bound states as each ETC boson
emits a top quark.  The end of the chain is a technipion which
materializes as a $W$ or $Z$.  The process $ gg \rarr t \bar t Z^0$
is expected to be an effective signature for ETC production
at the LHC up to
masses of about 1.5 TeV.  Though a 1.5 TeV $\ee$ collider will not
be able to reach the ETC pair resonance at these high energies,
one would expect to see the energy-dependent resonance enhancement
of $t\bar t$ and $t \bar t Z^0$ production.  At the highest energies,
one might see the associated production of $t$ quarks with
$\bar{\rm  ETC}$-technifermion bound states.

\REF\Chivukula{R. S. Chivukula, S. Selipsky, and E. H. Simmons,
        \sl Phys. Rev. Lett. \bf 69, \rm 575 (1992);
R. S. Chivukula, E. H. Simmons, and J. Terning,
      \sl Phys. Lett. \bf B311, \rm 383 (1994).}
\REF\CDF{F. Abe, \etal (CDF Collaboration), \sl Phys. Rev. Lett.
                   \bf 73, \rm 225 (1994).}
  More generally, the ETC renormalization of the top quark form factors
should produce effects of order  $(m_t/m_{\rm ETC})^2$ in the energy
dependence of $t$ quark pair production and may produce an effect of
order $(m_t/m_{\rm ETC})$ in the normalization of the $t \bar t Z^0$
form factor\refmark\Chivukula.
    If these  effects are of the
expected size, they  will require for their detection  the control
over the shape and absolute normalization of form factors
at the few percent level available
only at an $\ee$ collider.  (At this moment, there is much speculation
about larger effects, based on anomalies in the distribution of the
 first top quark events reported
by CDF\refmark\CDF.)    In any event, $\ee$ colliders offer many
handles for the separation and
measurement of the top quark production and decay
form factors, in particular, a large polarization asymmetry which
results from $\gamma$--$Z$ interference\refmark\MyFin.

\REF\DuganRan{M. J. Dugan and L. Randall, \sl Phys. Lett.
        \bf B264, \rm 154 (1991).}
\REF\GTe{E. Gates and J. Terning, \sl Phys. Rev. Lett.
             \bf 67, \rm 1840 (1991).}

  To conclude this section, I should point out that the version of
technicolor phenomenology that I have presented here is rather
conservative, to the extent that it does not take into account all
known difficulties of the technicolor scheme.  A techni-rho of the
size discussed above creates electroweak radiative corrections which
are now excluded by the precision $Z^0$ data at the 3 $\sigma$ level.
This problem can be cured by including  exotic pGb's or
Majorana fermions with mass about 100 GeV\refmark{\DuganRan,\GTe};
 these would be naturally
found at $\ee$ colliders.  The version of ETC discussed by Arnold
and Wendt has difficulty in producing a top quark mass above
100 GeV; the cure for this problem may dilute their
direct ETC signal, but it may also provide new corrections to the top
quark form factors.  It is not obvious whether the version of
technicolor chosen by Nature will include new high-energy signatures
or new corrections to the properties of the $W$ and $t$ which couple
strongly to this sector. We should be prepared to look for both of
these effects.

\newpart{Conclusions}

  In this lecture, I have tried to make a reasoned comparison of the
capabilities of the next generation of
$\ee$ and $pp$ colliders.  I have taken as my starting point the
idea that the goal of the next colliders to discover the mechanism
of electroweak symmetry breaking.  I have taken seriously that idea
that electroweak symmetry breaking has an explanation in physics,
and that this explanation requires a new sector of forces and
interactions.  To evaluate the relative power of electron and hadron
experiments, we must study models of symmetry breaking in their
entirety, looking at the variety of phenomena that each model
makes available and comparing the
very different signatures the colliders access
at comparable values of the underlying parameters.

 I have presented  a broad survey of this sort for
supersymmetry and technicolor models of electroweak symmetry
breaking.   I do not insist that one of these models
is chosen by Nature as the solution to the problem of electroweak
symmetry breaking.  Rather, I am attracted to these models because
they have been thoroughly analyzed in the literature, and that
analysis can form the groundwork for the broad-based comparison
that I have argued is required.
 Since supersymmetry models are well characterized
quantitatively, we were able to make detailed comparisons of processes
available to $\ee$ and $pp$ experiments.  In the case of technicolor
models, the comparisons we made were more rough and indicative.
However, the conclusions  derived from these two very different
models are surprisingly similar.

  Most strikingly, this comparison highlights the complementarity
of $\ee$ and $pp$ experimentation.  For almost every phenomenon
available to the LHC, we identified a signature of new physics
at an $\ee$ collider which addresses the same issue from a
different viewpoint.  At times, these signatures were quite
similar, as in the example of the resonant effect of the technirho
on $WZ$ and $WW$ production, but at other times they involved
completely different experiments, for example, the comparison of
chargino masses in $\ee$ to the gluino mass observed in $pp$
collisions.

  In both models, the values of $\ee$ center of mass energy at which
these complementary signatures become available was much lower
that the value inferred from the criterion of  `discovery reach'.
In fact, in both models, $\ee$ experiments at 500 GeV in the center
of mass already would have a significant impact.  In supersymmetric
models, an $\ee$ collider at this energy would already cover the
complete allowed
 mass region for the lightest Higgs boson, allowing the detailed
characterization of this state, including the measurement of
decay branching ratios.  It would also cover the expected region
for pair-production of the lightest chargino.  In technicolor models,
a 500 GeV $\ee$ collider would allow precision measurement of the
top quark form factors, giving a window into ETC physics as likely
as any we discussed.  Though technicolor models seemed to put more
of a premium on experiments at high energy, we saw that,
at about 1 TeV in the supersymmetry case and at
about 1.5 TeV in the technicolor case, an $\ee$ collider
 would surpass the LHC in
providing experimental information on the new sector of interactions.

  In addition, we saw several examples in which information from
$\ee$ colliders is needed to fully interpret the experimental
results obtained from $pp$ colliders.  We saw this connection most
clearly in the case of supersymmetry, where the theory makes
detailed
  predictions for the properties of the Higgs boson and the
gluino which depend on a complex of model parameters, and where
$\ee$ colliders provide a systematic program for determining
those parameters.
It is quite likely that, if a quantitative
technicolor model of the top quark mass could be found,
experiments on large transverse momentum top production in $pp$
colliders would have the same relation to $\ee$
precision measurements
of the top quark properties at threshold.

  From all of these considerations, I conclude that an $\ee$ linear
collider, scalable in energy but beginning at 500 GeV in the center
of mass, will play an essential role in the experimental solution of
the problem of electroweak symmetry breaking.  It follows that we
should plan to make such a facility available
{\it simultaneously} with the LHC.

  The creation of two new accelerator facilities, each with a cost
in the billions of dollars, is not a simple task.  For most of the
lifetime of our field, we have justified the construction of new
facilities through international or inter-regional competition.
Almost twenty years ago, Prof. Yoshio Yamaguchi, the long-time
director of this series of workshops, introduced a different
vision of high-energy physics, based on global  cooperation
in international facilities.   When I first visited Japan in 1985,
Prof. Yamaguchi's vision seemed hopelessly idealistic, at a time
when the SSC and the LHC were being pursued as
competitive regional projects which would exhaust our global
resources.
  Today, it is an idea whose time
has finally come.   I hope that Prof. Yamaguchi's vision can be
combined with a clear appreciation of the physics issues
of the coming generation of colliders to provide the tools we will
need to understand the next level of the fundamental interactions.

\ack

I am grateful to Prof. Sakue Yamada for the invitation  to
      present this lecture, and to  Tim Barklow,
       David Burke, Jonathan Feng,
         Howard Haber, Gordon Kane,
       Hitoshi Murayama, Thomas Rizzo,
       Xerxes Tata, and many other colleagues
        at SLAC and elsewhere for discussion of the issues
              raised here.

\endpage
\refout
\endpage
\figout
\endpage

\vsize=8.5in
\input TABLES
\long\def\para#1{
   {%
   \vtop{%
         \hsize=\parasize%
         \baselineskip18pt%
         \lineskip1pt%
         \lineskiplimit1pt%
         \noindent #1%
         \vrule width0pt depth6pt%
      }%
   }%
}%
\nopagenumbers
\parasize=2.5in
\overfullrule0pt
\def\epem{{e^+e^-}}
\def\noET{\not{\hbox{\kern-4pt $E_T$}}}

\centerline{\seventeenrm Table 1: Supersymmetry}
\bigskip
\settabs\+ &\hbox{\hskip.80in}\quad\qquad
           &\hbox{\hskip2in}\qquad\qquad
           &\hbox{\hskip2in}\cr
\+&{\bf Issues}  &\qquad\qquad $\bold{\epem}$
                 &\qquad\quad $\bold{pp}$ {\bf (LHC)} \cr
\+&       &    & \cr
\+&{\bf Higgs:}\hbox{\hskip 2.2in} $m(h^0) < 150 \ GeV$ && \cr
\+&       &    & \cr
\+&$h^0 $     &\para{visible in $\epem \rarrow Z^0h^0$ \brk
            BR measurements  for $b\bar b,\tau^+\tau^-,\brk
            c\bar c,WW^*$, $\Gamma(h^0\rarrow \gamma\gamma)$}
           &\para{visible in $h^0\rarrow \gamma\gamma$ in most of
            \brk parameter space  \brk \hbox{
            $BR(h^0\rarrow \gamma\gamma)
            \Gamma(h^0\rarrow gg)$}}\cr
\+&       &    & \cr
\+&$H^0,A^0$
&\para{visible in $\epem\rarrow H^0A^0$ \brk up to threshold}
          &\para{visible in $H^0,A^0\rarrow\tau^+\tau^-$ if \brk
            $\tan\beta\gsim 10$}\cr
\+&       &    & \cr
\+& $H^+$
          &\para{visible in  $\epem\rarrow H^+H^-$ \brk up to threshold}
          &\para{visible in  $t\rarrow bH^+$ if \brk $m_H\lsim
            120\ GeV$} \cr
\+&       &    & \cr
\+&{\bf Superpartners:}       &    & \cr
\+& gauginos \hbox{\hskip 2in}
 $m(\widetilde\chi^+)< {1\over 4}\, m(\widetilde g)$ \cr
\+&       &    & \cr
\+&
          &\para{$\widetilde \chi^-\widetilde\chi^+$ observable up to
            threshold  \brk
            $\Delta m_\chi/m_\chi\sim 2\%$
            for $\widetilde\chi^+,\widetilde\chi^0$,  \brk
            $m_0,m_{1/2},\mu \rarrow$ tests of unifi\-ca\-tion,
            $\widetilde\chi_1^+,\widetilde\chi_2^+\rarrow$
            tests of  super\-symmetry}
          &\para{$\widetilde g$ observable up to 1.6~TeV in $\noET$,
            $\ell^\pm\ell^\pm,\ell Z^0$,
            $\Delta m\widetilde g/ m\widetilde g\sim 10\%$
            under model-dependent hypotheses, \brk
            $\widetilde \chi^+$ observable up to 150~GeV in
              \brk multileptons } \cr
\+&       &    &\cr
\+&       &    & \cr
\+&\para{squarks,\brk sleptons}
&\para{visible in $\epem\rarrow\widetilde\ell^+\widetilde\ell^-$ \brk up
      to threshold  \brk ${\cal O}(1)$ polariz.  effects,  \brk
      ${\Delta m_\ell/
      m_\ell}\sim 1\%$}
          &\para{visible in $gg\rarrow \widetilde q\widetilde q$
            unless buried \brk by the $\widetilde g\widetilde g$
                 signal}\cr
\endpage
\long\def\para#1{
   {%
   \vtop{%
         \hsize=\parasize%
         \baselineskip18pt%
         \lineskip1pt%
         \lineskiplimit1pt%
         \noindent #1%
         \vrule width0pt depth6pt%
      }%
   }%
}%
\nopagenumbers
\parasize=2in
\long\def\wpara#1{
   {%
   \vtop{%
         \hsize=4.0in%
         \baselineskip18pt%
         \lineskip1pt%
         \lineskiplimit1pt%
         \noindent #1%
         \vrule width0pt depth8pt%
      }%
   }%
}%
\overfullrule0pt
\def\epem{{e^+e^-}}

\centerline{\seventeenrm Table 2: Technicolor}
\bigskip
\settabs\+ &\hbox{\hskip1in}\quad\qquad
           &\hbox{\hskip2in}\qquad\qquad
           &\hbox{\hskip2in}\cr
\+&{\bf Issues}  &\qquad \qquad $\bold{\epem}$
                 &\qquad \quad $\bold{pp}$ {\bf (LHC)} \cr
\+&        &   &  \cr
\+&Techni-rho
           &\para{visible above 4 TeV as an enhancement of $\epem
                  \rarrow W^+W^-$}
          & \para{visible up to 2 TeV in \brk $(T\rho)\rarrow WZ$} \cr
\+&       &    & \cr
\+&        &   &  \cr
\+&top    &\para{ETC effect on $t$ production form factors}
          &\para{ETC $\rarrow t\bar t + W\ {\rm or}\ Z$} \cr
\+&       &    & \cr
\+&        &   &  \cr
\+&\para{Pseudo-Goldstone \brk bosons} &
  \wpara{PGB's may be colored, but \brk
  {\it must} have electroweak charge.}\cr
\+& \qquad
          &\para{visible in $\epem \rarrow P\bar P$,
                 up to threshold}
          &\para{visible in $gg\rarrow P\bar P$ in $\ell+$ jet modes or
in $P\rarrow t \bar t$ if $P$ very light} \cr
\+&       &    &\cr
\+&        &   &  \cr
\+&\para{$S<0$}
          &\para{spectrum must include light PGB's or light Majorana
                 fermions}\cr
\endpage
\end